\documentclass[aps,twocolumn,pra,superscript,floatfix,superscriptaddress,showpacs,footinbib,reprint]{revtex4-2}
\DeclareUnicodeCharacter{2212}{\textminus}
\usepackage[plainpages=false,pdfpagelabels,colorlinks=true,linkcolor=red,urlcolor=blue,
citecolor=blue,pdftitle={kagome},pdfauthor={},pdfdisplaydoctitle=true]{hyperref}
\usepackage{tikz}
\usepackage{graphicx}
\usepackage{titlesec} 
\usepackage{float}
\usepackage{amsthm}
\usepackage{amsmath}
\usepackage{amsfonts}
\usepackage{amssymb}
\usepackage{array}
\usepackage{hhline}
\usepackage{nicefrac}
\usepackage{color}
\usepackage[normalem]{ulem} 
\usepackage{subfigure}
\usepackage{dsfont}
\usepackage{txfonts}
\usepackage{wasysym}
\usepackage{multirow}
\usepackage{sidecap}
\usepackage{xcolor,cancel}
\usepackage{braket}
\usepackage{acronym}

\begin{document}

\title{Exploring the Kondo Effect in Strained Kagome Nanoribbons}

\author{P.~A.~Almeida}
\email{passis@if.usp.br}
\affiliation{Instituto de Física, Universidade de São Paulo, C.P. 66318, 05315--970 São Paulo, SP, Brazil}
\affiliation{Instituto de F\'isica, Universidade Federal de Uberl\^andia, 
Uberl\^andia, Minas Gerais 38400-902, Brazil}
\affiliation{Department of Physics and Astronomy and Nanoscale and Quantum Phenomena Institute,
Ohio University, Athens, Ohio 45701-2979, USA}

\author{G.~B.~Martins}
\email{gbmartins@ufu.br}
\affiliation{Instituto de F\'isica, Universidade Federal de Uberl\^andia, 
Uberl\^andia, Minas Gerais 38400-902, Brazil}

\author{S.~E.~Ulloa}
\email{ulloa@ohio.edu}
\affiliation{Department of Physics and Astronomy and Nanoscale and Quantum Phenomena Institute,
Ohio University, Athens, Ohio 45701-2979, USA}

\date{\today}
\begin{abstract}
Metallic kagome systems have attracted considerable interest in recent years, as they provide a rich platform for studying phenomena associated with their distinctive band structure. The coexistence of bands with Dirac points similar to those in graphene, along with a completely flat band, makes this an ideal structure for investigating how lattice symmetries may protect topological and many-body correlation effects. 
Since applied strain can break lattice symmetries and modify the electronic structure, understanding how strain influences phenomena such as the Kondo effect in kagome materials may provide essential insights into correlated-electron behavior.
We employ the single-impurity Anderson model and the numerical renormalization group to analyze the Kondo effect in kagome zigzag  nanoribbons under uniaxial strain. 
We find that strain manipulation enables precise control over the strength of the Kondo effect on an impurity hybridized on the ribbon with different coordination environments,
and that symmetric local environments may result in strong suppression of the effective hybridization due to orbital interference.  We find that the specific location of the impurity on the ribbon, especially when the Fermi energy lies close to weakly dispersive edge states, can lead to significant changes in the characteristic Kondo temperature. Such sensitivity may be used to provide unique 
information on the local density of states of a system.  
These results demonstrate that strain is a powerful tuning parameter in kagome nanoribbons, strongly modifying the screening of magnetic impurities.  
\end{abstract}
\maketitle

\section{Introduction}

The Kondo effect is likely one of the most extensively studied phenomena in strongly correlated electron systems \cite{kouwenhoven2001revival, pustilnik2004kondo,goldhaber1998kondo, hewson1993kondo}. Elucidated by J. Kondo in 1964 \cite{kondo1964resistance}, it provided a compelling explanation for the well-known low temperature resistance minimum in metallic alloys containing diluted magnetic impurities \cite{de1934electrical}. The presence of conduction electrons leads to the dynamical screening of magnetic impurities by the formation of a many-body spin singlet ground-state \cite{jarillo2005orbital, hewson1993kondo, krishna1980renormalization, nozieres1980kondo}. The Kondo effect has been also realized in a variety of other systems, including heavy fermion metals \cite{coleman2010frustration, checkelsky2024flat}, mixed-valence compounds \cite{barzykin2005competition}, quantum dots \cite{van2002two, pustilnik2000quantum, cronenwett1998tunable, sasaki2000kondo}, adatoms and molecules on metallic surfaces \cite{madhavan,schneider,sawkondo}, and topological insulators \cite{tran2012phase, tran2010probing, wang2013suppression}. As the onset of the Kondo screening and associated energy scale (the Kondo temperature) depend on the density of carriers in the host and the effective exchange coupling strength of the impurity, it is of interest to be able to modify these quantities in a given system.  Doping or gating, which produce shifts in chemical potential, may achieve that objective, while strain applied to the host material can modify the local electronic states, affecting the Kondo behavior \cite{Ruthenium}.  
In this work, we discuss our findings on the application of uniform uniaxial strain on a kagome system. We find that this approach can provide an effective tuning parameter for modifying Kondo physics, as the strain breaks lattice symmetries and can strongly modify the local density of states (DOS) near the Fermi energy ($E_F$).

The investigation of the Kondo effect has moved beyond conventional metals, characterized by a flat DOS near the Fermi energy \cite{beri2012topological, sasaki2000kondo, grobis2007kondo, almeida2025identifying}, extending into systems with more complex electronic structures. Kondo physics in two-dimensional materials \cite{granet2018tuning}, such as monolayer graphene, may exhibit suppression of the Kondo temperature due to the vanishing DOS of the Dirac cone at low energies \cite{sengupta2008tuning, li2013kondo,chen-fuhrer2011,wehling2010orbitally, zhu2011magnetic}. Suppression of the Kondo screening in pseudogapped systems has been found to exhibit interesting spectral features \cite{KevinQimiao}, and can be implemented in coupled quantum dot systems \cite{LuisDQD}.  In contrast, the emergence of flat-bands or pronounced van Hove singularities (VHS) in various material platforms \cite{hu2005effects,PatriciaCNT}, including twisted bilayer graphene \cite{shankar2023kondo}, heterobilayer transition metal dichalcogenides \cite{xie2024kondo}, and carbon nanotubes \cite{PatriciaCNT} results in remarkable Kondo behavior as the DOS is enhanced, which may lead to intense competition between the Kondo screening of local moments and alternative correlation-induced ground states.

Among the lattice structure geometries known for their interesting magnetic and electronic properties, the kagome lattice frequently underlies the behavior of both insulators and metals of importance \cite{uemura1994spin, chalker1992hidden, kang2020topological, hagemann2001geometric, nocera2004spin}. Kagome materials, featuring flat bands, nearby dispersive Dirac points, and van Hove singularities, have been linked to the emergence of exotic phases of matter  \cite{ye2018massive, ko2009doped, yan2011spin}, and stand out as prime platforms for probing phenomena driven by strong electron correlations. While the Kondo effect has been examined in flat-band systems \cite{tran2018molecular, ukpong2023kondo, tokuda2022kondo}, it is only recently being explored within kagome lattices \cite{checkelsky2024flat}. Further investigation is required to elucidate how lattice structure and electron correlations interplay to produce Kondo physics in these materials.

Straintronics has been recognized as a powerful approach for modulating the electronic properties of quantum materials, where mechanical deformation can fine-tune their electrical behavior \cite{bukharaev2018straintronics, si2016strain}. For example, uniaxial deformation has been studied to control magnetic frustration in a kagome lattice, noticeably increasing its Neel temperature \cite{wang2023controlled}. Strain has been shown to significantly affect van Hove singularities (VHSs), which in turn influence the superconductivity seen in the kagome metal CsV$_3$Sb$_5$ \cite{lin2024uniaxial}, possibly enabling the coexistence of superconducting and CDW phases \cite{mojarro2024tuning}. Strain engineering may thus provide a unique approach for the systematic control of quantum materials, opening new possibilities for their study and applications.

In this work, we present a study of magnetic impurities incorporated in strained kagome systems and the resulting Kondo physics. We study strain in kagome nanoribbons, which offer a convenient means to precisely tune the Kondo screening provided by the host carriers.
The coordination environment of a magnetic impurity in a lattice (e.g., in a hollow triangle or on a bridge site), and its proximity to the edge of a finite ribbon, are expected to significantly influence the effective exchange environment through the local DOS. We focus on the role of strain in modifying the local host environment, enhancing control over Kondo screening and Kondo temperatures, and enabling new quantum states and straintronic applications. 

We find that the spectral function and associated Kondo temperature scale depend critically on the coordination environment and the location of the impurity. The screening is significantly weakened for symmetric hollow sites due to destructive interference among the various hybridization pathways.  In contrast, the on-top coordination results in stronger coupling and variation of Kondo temperature with chemical potential.   Interestingly, we find that for impurities near the edges of a zigzag nanoribbon that host structurally protected states, the Kondo scale is typically much larger than for the same coordination in the middle of the ribbon.  This location-specific dependence arises from the larger local density of states near the end of the ribbon when the Fermi energy approaches edge-protected states. Moreover, we find that even a weak strain can change screening and Kondo temperatures by large factors, especially when the Fermi energy is close to van Hove singularities or the flatband in the kagome system.  This sensitivity to strain on a many-body effect suggests the possible use of controlled mechanical perturbations to modify the ground state of interacting systems, and the possibility to study its competition with topological features of the single-particle spectrum.   

The remainder of the paper is organized as follows. In Sec.\ II, we introduce a tight-binding model to describe a kagome nanoribbon and compare its electronic behavior under strain to that of the corresponding two-dimensional lattice. 
This section also includes a description of the Anderson impurity model, and the theoretical approach we use to calculate the many-body screening properties of this unique host. In Sec.\ III  we present our key results on the impact of different strain configurations and impurity locations on the Kondo temperature as well as the spectral properties and thermodynamic features of the impurity.  Sec.\ IV provides a discussion of the results and possible experimental consequences, as well as our overall conclusions.

\section{Model and Methods}

\subsection{Tight-binding 2D kagome lattice}

\begin{figure}[ht]
    \centering
     \includegraphics[width=1.0\columnwidth]{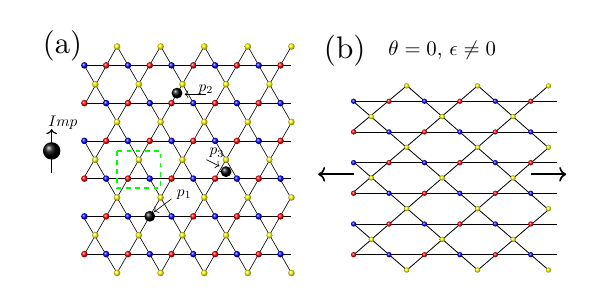} \vspace{-4ex}
     
     \includegraphics[width=1.0\columnwidth]{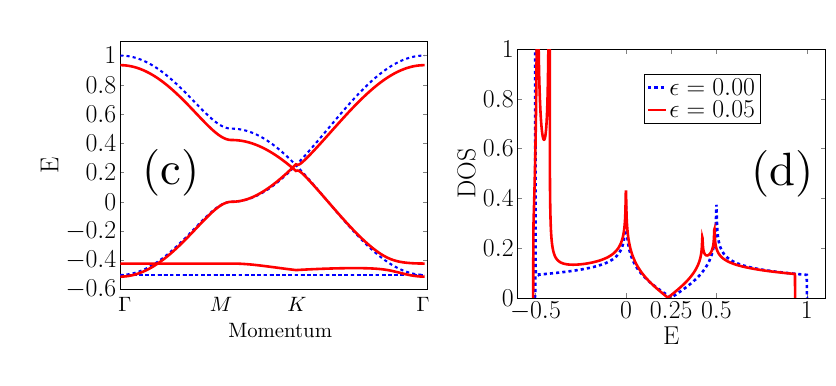} 
    \caption{(a) Schematic representation of a kagome ribbon with zigzag edges at top/bottom, and with periodic boundary conditions 
    along the horizontal direction. The green dashed rectangle indicates a unit cell of the kagome lattice, composed of three sites. Black spheres stand for magnetic impurities in different configurations, $p_1$, $p_2$, and $p_3$, identified as on top, bridge, and hollow sites, respectively. 
    (b) Ribbon after strain is applied, illustrating the resulting deformation of the system, here at an angle $\theta = 0$ (indicated by black arrows). (c) Band structure of a 2D kagome lattice along the high-symmetry path $\Gamma$-$M$-$K$-$\Gamma$. The blue dashed lines show the unstrained case, while the red solid lines correspond to the strained system. The deformation leads to notable modifications, including shifting Dirac points near $K$, a shift in the upper band near the $M$ point, and changes in the flat-band region.  (d) Corresponding DOS curves for panel (c).  Notice that strain reduces overall bandwidth, shifting DOS especially near the top edge. A structure near the higher VHS, along with drastic changes in the low-energy edge region, is also visible.}
    \label{fig1} 
\end{figure}

The kagome lattice takes its name from its resemblance to a traditional Japanese basket-weaving pattern \cite{mekata2003kagome}. It exhibits an interconnected triangular honeycomb arrangement with atomic coordination number of four. The kagome is defined on a triangular lattice with three distinct sites per unit cell [see dashed green rectangle in Fig.\ \ref{fig1}(a)], and nearest-neighbor vectors given by $\Vec{d}_{13} = (1,\sqrt{3})/2$, $\Vec{d}_{12} = (1,0)$, and $\Vec{d}_{23} = (-1,\sqrt{3})/2$. In real space, the tight-binding kagome Hamiltonian can be expressed as 
\begin{equation}
    H_{\text{kagome}} = \sum_{\langle i,j \rangle \sigma} t \left( c_{i \sigma}^{\dagger} c_{j \sigma} + h.c. \right),
    \label{eq1}
\end{equation}

\noindent
where $c_{j \sigma}^{\dagger}$ and $c_{j \sigma}$ are the electron creation and annihilation operators at the $j$ lattice site, with spin $\sigma=\uparrow, \downarrow$, and the sum considers equal hopping amplitude $t$ between nearest-neighbors in the unstrained system. The corresponding Hamiltonian in reciprocal space, $\tilde{H}_{\text{kagome}}(\vec{k})$, then reads
\begin{equation}
    \tilde{H}_{\text{kagome}}(\vec{k}) = 2t
\begin{pmatrix}
0 & \cos \gamma_{12} & \cos \gamma_{13} \\
\cos \gamma_{21} & 0 & \cos \gamma_{23}\\
\cos \gamma_{31} & \cos \gamma_{32} & 0
\end{pmatrix},
\label{eq2}
\end{equation}
using a basis $(c_A^\dagger,c_B^\dagger,c_C^\dagger)$ to denote the three sites within the unit cell (shown as red, blue, and yellow in Fig.\ \ref{fig1}), and 
where $\gamma_{ij} = \vec{k} \cdot \vec{d}_{ij} $. Diagonalizing Eq.\ (\ref{eq2}) yields the energy spectrum, which consists of three bands, given by
\begin{equation}
    E_{1,2,3}(k) =
\begin{cases} 
-2t,\\ 
t \left( 1 \pm \sqrt{4 (\cos^2 \gamma_{12}+ \cos^2 \gamma_{13} + \cos^2 \gamma_{23}) - 3} \right). \\
\end{cases}
\end{equation}
\noindent

We find it convenient to rescale the bandwidth to contain the entire spectrum within the region $(-1,1)$, by setting $t=1/4$ from now on.
The two dispersive bands, $E_{2}(k)$ and $E_{3}(k)$ [dashed blue curves in Fig.\ \ref{fig1}(c)], meet at the $K$ corners of the Brillouin zone (BZ)  ($E=0.25$), creating Dirac cones with linear dispersion, similar to those in graphene \cite{katsnelson2007graphene}. Most peculiar, the system also has a completely flatband at $E_1=-0.5$ (touching the dispersive band $E_3$ at the $\Gamma$ point). The coexistence of these distinctive features makes the kagome lattice an ideal platform for exploring exotic phases of matter. Figure \ref{fig1}(d) shows the corresponding total DOS (dashed blue curve): in addition to the Dirac cone region with vanishing DOS at $E=0.25$, a flat-band ($E=-0.5$) and two van Hove singularities ($E=0$ and $E=0.5$) are clearly observed.

\subsection{Uniaxial strain in 2D kagome}
Applying hydrostatic pressure or uniaxial strain to a crystalline system modifies its physical properties by directly altering interatomic distances, resulting in strong modification of hopping amplitudes. The change in atomic positions due to a uniaxial deformation is given by $\vec{r}_{\text{new}} = (\mathbb{I} + \bar{\epsilon}) \vec{r}_{\text{old}}$, changing the hopping parameter to $t_{\text{new}} = t e^{-\beta \, |\Delta \vec{r}\,|}$, where $\Delta \vec{r} = \bar{\epsilon}\cdot \vec{r}_{\text{old}}$ \cite{pereira2009tight}, and $ \bar{\epsilon}$ is the strain tensor, given by 
\begin{equation}
\bar{\epsilon} =\epsilon
\begin{pmatrix}
\cos^2 \theta - \nu \sin^2 \theta & (1 + \nu) \cos \theta \sin \theta \\
(1 + \nu) \cos \theta \sin \theta & \sin^2 \theta - \nu \cos^2 \theta
\end{pmatrix}.
\end{equation}
In the above, $\beta \approx 3$ is the Gr\"uneisen parameter and $\nu$ is the Poisson ratio (set to 0.165) \cite{mojarro2024tuning}. The strain of magnitude $\epsilon$ is applied at an angle $\theta$, defined relative to the horizontal direction ($x$-axis) in Fig.\ \ref{fig1}(b)].

Let us first examine how the system behaves under strain by focusing on the band structure. 
Figure \ref{fig1}(c) shows the characteristic dispersion along high symmetry directions in the Brillouin zone, 
solid red lines showing a system under strain (dashed blue lines, unstrained case); the corresponding DOS are shown in Fig.\ \ref{fig1}(d). The strain modifies all energy bands, particularly in the region near the upper VHS ($E \simeq 0.5$) and around the flat-band ($E \simeq -0.5$); it also reduces the overall bandwidth and shifts the Dirac cone location  
 away from the $K$ point, in a direction that depends on the strain angle and magnitude \cite{mojarro2Dmats}. The breaking of lattice symmetries introduced by the strain also produces a finite bandwith of the flat-band, as well as the creation of additional Dirac points near the spectrum edge \cite{mojarro2Dmats}. These features can be seen near the bottom of the spectrum in Fig.\ \ref{fig1}(c) and (d).

\subsection{Electronic structure of zigzag kagome nanoribbons}

\begin{figure}[ht]
  \centering
  \includegraphics[width=1.0\columnwidth]{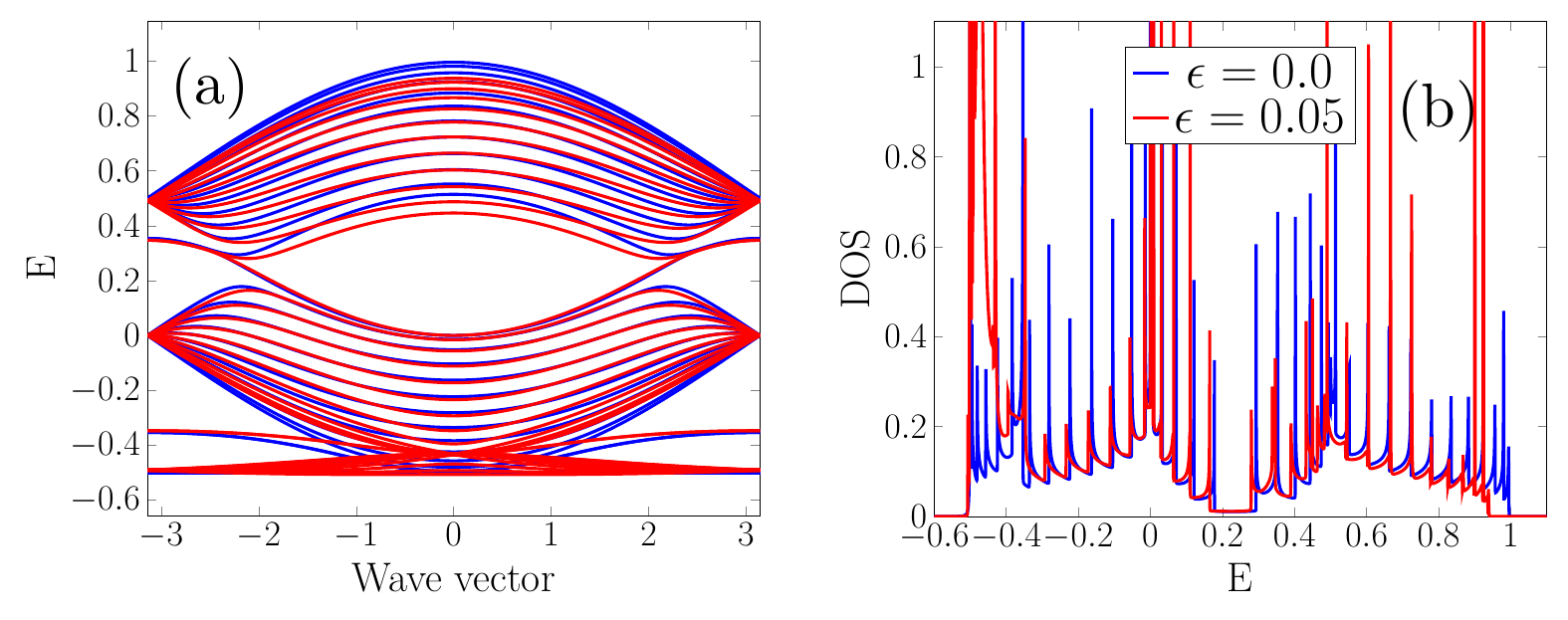}
\caption{(a) Subband structure for a ZKR, without (blue) and with (red) applied strain ($\epsilon=0.05$). Strain modifies subband dispersion, particularly near the top of the range ($E \simeq 1$), the low-energy region of the flat-band ($E\simeq -0.5$), and the upper VHS ($E\simeq 0.5$). (b) Corresponding DOS highlights strain-induced changes due to subband shift and mixing, especially in the regions mentioned in (a).}
  \label{fig2}
\end{figure}

Figure~\ref{fig2} shows the energy spectrum of a zigzag kagome ribbon (ZKR) of finite width, a quasi-one-dimensional system obtained by cutting a 2D kagome lattice--Fig.\ \ref{fig1}(a) shows a ZKR with upper and lower edges terminated by yellow atoms, while having periodic boundary conditions along the horizontal direction ($x$-axis). This geometry results in a collection of 1D subbands, as shown in Fig.\ \ref{fig2}(a); as before, the red curves represent the system under $\epsilon=0.05$ uniaxial strain ($\theta = 40^\circ$), while blue curves correspond to the unstrained case ($\epsilon=0$). Figure \ref{fig2}(b) 
shows the corresponding DOS for panel (a), showing 1D-like sharp features at the bottom and top of each subband in the ribbon. 
As in the 2D lattice, applying uniaxial strain induces significant modifications in the ZKR electronic states, including the distortion of the flatband and higher VHS. 
Significantly, the zigzag edges in the ZKR support weakly dispersing states with energies $E \simeq \pm 0.37$ at the edge of the BZ, Fig.\ \ref{fig2}(a).  
These states extend along the edge but not into the middle of the ribbon.
In the presence of spin-orbit interactions, they have been shown to evolve into spin-valley polarized edge states \cite{sun2022spin}.  Although not topologically protected in our system, 
the spatial localization of these states near the ribbon edges has
significant consequences in the resulting screening of a magnetic impurity, as we will discuss below.

\begin{figure}[ht]
    \centering
    \includegraphics[]{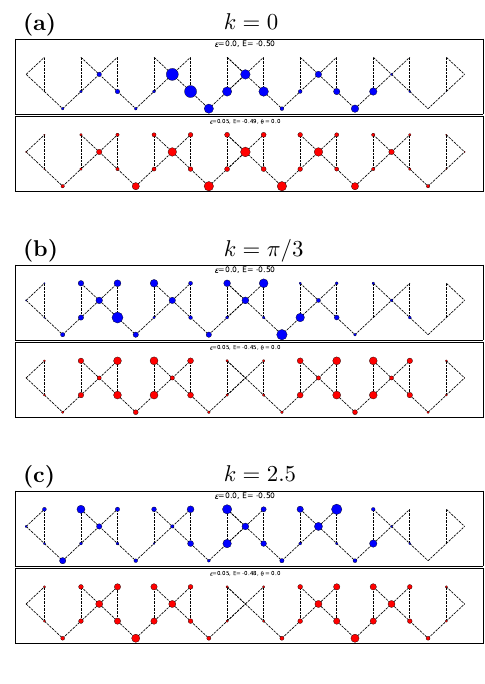}
\caption{Spatial distribution for different ZKR eigenstates. The zigzag edges are on the left/right of the figures. The three sets of 
panels (for different wave vector $k$, as indicated) show results without 
(blue) and with (red) uniaxial strain ($\epsilon =5\%$, $\theta=0$), respectively. Energy value for each pair are (a) $E=-0.5$ (without) and $E=-0.49$ (with strain) for $k=0$; (b) $E=-0.5$ and $E=-0.45$ for $k=\pi/3$; (c) $E=-0.50$ and $E=-0.48$ for $k=2.5$. Spatial weights shift across the ribbon as strain is applied, including the weights on the zigzag edges. } 
  \label{fig3}
\end{figure}

Figure~\ref{fig3} shows a few characteristic states in a ZKR.\@  For convenience, the ribbon has been rotated with respect to 
Fig.\ \ref{fig1}(a), i.e., the zigzag edges are displayed vertically here, while the ribbon unit cell is repeated periodically in the vertical direction. A generic state in the system is given by
$|\psi \rangle = \sum_i \, \psi_i |i \rangle$, where the sum is over all the sites in the ribbon, and $\psi_i$ depends on energy and $k$ value in the 1D BZ.  Figure~\ref{fig3} shows maps of $|\psi_i|^2$, for different values of $k$ and $E$, with and without strain.  As the strain shifts the energy of a given subband $k$-state, the corresponding wavefunction is redistributed differently across the system, typically shifting the weights towards the middle of the ribbon, while in others it accentuates the weight on the zigzag edges. This redistribution suggests that, as the magnetic impurity is incorporated at different locations and with different local geometries or environments (the different $p_i$'s in Fig.\ \ref{fig1}(a)), the associated hybridization function will change with strain and result in amplitudes that depend on chemical potential and local DOS.

\subsection{Anderson Impurity Model}
To investigate the physical properties of a magnetic impurity in a kagome ribbon, the single-impurity Anderson model (SIAM) is an essential tool. Formulated by P.~W. Anderson  \cite{anderson1961localized}, this model is fundamental in the analysis of strongly correlated electron systems.
The Hamiltonian is expressed as the sum of three terms:
\begin{equation}
    H = H_{\text{imp}} + H_{\text{ribbon}} + H_{\text{Hyb}}.
\end{equation}
The impurity Hamiltonian, $H_{\text{imp}}$, can be written as
\begin{equation} 
    H_{\text{imp}} = \sum_{\sigma} \epsilon_f f_{\sigma}^{\dagger}f_{\sigma} + U n_{f \uparrow} n_{f \downarrow}.
\end{equation}

\noindent
Here, the operators $f_{\sigma}$ and $f_{\sigma}^{\dagger}$ correspond to annihilation and creation operators for an electron at the impurity with spin $\sigma$, and $n_{f\sigma}=f_{\sigma}^{\dagger}f_{\sigma}$. $\epsilon_f$ defines the orbital energy at the impurity, while $ U$ accounts for the on-site Coulomb interaction, which energetically penalizes double occupancy. 

The Hamiltonian $H_{\text{ribbon}}$ models the host system, including the chemical potential $\mu$ that controls the ribbon occupation (or filling factor), 
\begin{equation}
    H_{\text{ribbon}} = \sum_{i\sigma} \mu c_{i\sigma}^{\dagger} c_{i\sigma} + H_{\text{kagome}},
\end{equation}
with the appropriate boundary conditions to describe the ribbon.
Finally, the interaction between the impurity and the host system is captured by $H_{\text{Hyb}}$ , 
\begin{equation}
    H_{\text{Hyb}} = \sum_{i, \sigma} (V_i \, c^{\dagger}_{i\sigma} f_{\sigma} + h.c.).
\end{equation}
The local hybridization amplitude $V_i$ quantifies the interaction between the impurity and the lattice, defining the geometry under which the magnetic impurity is incorporated into the kagome lattice. We consider three possible natural configurations--see Fig.\ \ref{fig1}(a): a {\em hollow site} $p_3$ configuration, where the impurity couples equally to three sites in a unit cell ($V_1=V_2=V_3$); a {\em bridge} $p_2$ configuration, where the impurity couples to only two nearest-neighbor sites ($V_1=V_2, V_3=0$); and an {\em on top} $p_1$ configuration that couples to a single site.

We should notice that in the well-known SIAM for a host with a featureless DOS, the Kondo screening is determined by the effective hybridization strength $\Gamma=V^2\rho(E_F)/\pi$, where $\rho(E_F)$ is the host DOS at the Fermi level and $V$ is the local coupling \cite{hewson1993kondo}. In contrast, the effective hybridization strength in the ZKR for the different impurity configurations changes depending on the location of the impurity in the ribbon, Fermi energy, and applied strain. In our calculations, we determine the effective hybridization function
\begin{equation}
    \Delta(\omega) = \sum_{jk} |V_j|^2 |\psi_j|^2 \, \delta(\omega-E_k),
\end{equation}
through the use of recursive Green's functions (GFs). The ribbon GF is calculated by assuming the system is composed of three kagome regions: the left contact, the central region containing the impurity, and the right contact. The leads are modeled as semi-infinite, while the central region retarded GF is given by
\begin{equation}
     G_C(\omega) = \big(\omega + i\eta - H_C - \Sigma_L - \Sigma_R\big)^{-1}.
\end{equation}
Here $H_C$ describes the central region Hamiltonian and $\Sigma_{L/R}$ represent the self-energies of the contacts, calculated as
\begin{equation}
    \Sigma_l = H_{lC}^\dagger g_l H_{lC}, 
\end{equation}
\noindent
with the lead's GF $g_l$ calculated iteratively for $l=L/R$, and $H_{lC}$ couples each lead to the central region. The impurity hybridization is given by $\Delta(\omega) = \mathrm{Im}\left[ G_C^{-1}(\omega) \right]_{NN}$, where $N$ designates the impurity site
 \cite{diniz2018kondo}.

The strong variation in the local DOS as different parameters change can be incorporated by employing the numerical renormalization group (NRG) method, a reliable approach for studying magnetic impurity screening in a variety of systems \cite{Rnrg-review}. Our calculations are implemented using the Ljubljana open-source code \cite{zitko}, with a discretization parameter $\Lambda = 2.0$ to ensure accurate resolution while retaining at least 10,000 states at each iteration. We also use the $z$-trick \cite{campo2005alternative}, with values of $z$ ranging from 0.0625 to 1.0 (i.e., $N_z = 16$), to minimize artifacts in the computed physical quantities. Furthermore, thermodynamic quantities are determined using the conventional single-shell approximation, whereas dynamical quantities, including the spectral function, were computed through the density matrix NRG approximation \cite{hofstetter2000generalized}. For simplicity, we consider a particle-hole symmetric configuration for the impurity, with
$U = 0.01$ and $\epsilon_f = -U/2$, for all calculations. 

\section{Results: Impact of uniaxial strain on the Kondo effect for the infinite system}

As discussed above, the plots for different eigenstates in Fig.\ \ref{fig3} illustrate the effect that strain has on redistributing the wavefunction weights on the lattice. 
This effect depends, of course, on the energy and characteristics of the state, as well as on the strain direction.
As the impurity is incorporated onto the lattice in a given configuration $p_j$, its coupling amplitude is modified by strain, since the latter changes the local DOS. It is that sensitivity to strain that allows one to control the Kondo screening behavior provided by the host, even at a fixed chemical potential. This also provides a sensitive response to the impurity's location on the nanoribbon, as the hybridization function depends on it.

In what follows, we analyze results for different impurity configurations on various locations of a ZKR. We show that impurity screening properties depend on its position on the ribbon. Moreover, variations in parameters such as strain direction, strength, and chemical potential lead to significant changes in both the dynamic and static characteristics of the system.

\begin{widetext}

\begin{figure}[ht]
    \centering
     \includegraphics[width=1.0\columnwidth]{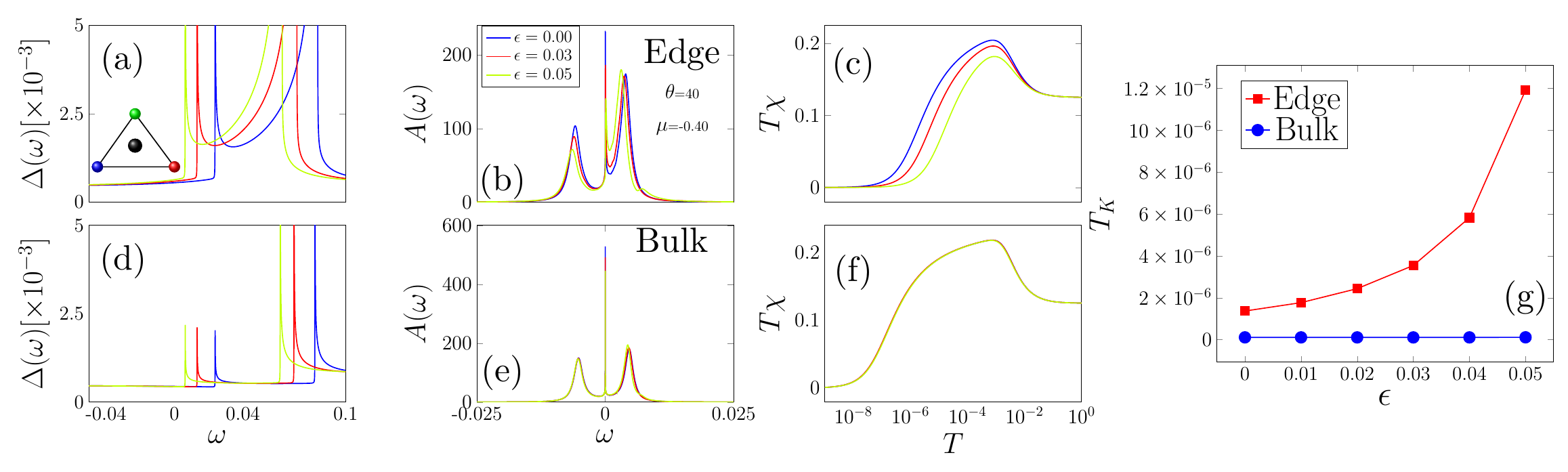}
    \caption{Results for an impurity in hollow site configurations ${p}_3$ located at two different locations on the ZKR. Panels in top row (a-c) show results for an impurity on one edge of the ribbon (`Edge'), while the bottom row (d-f) shows results for an impurity in the middle of the ribbon (`Bulk'). In both cases, $\mu =-0.40$ and $\theta=40^\circ$. Panels (a) and (d) show the corresponding hybridization function $\Delta(\omega)$ for different values of strain; notice that $\Delta(\omega)$ shifts to lower energies as strain increases. Peaks near the Fermi energy ($\omega \simeq 0$) are associated with states residing mostly on the ribbon edge; higher energy features correspond to states extending throughout the ZKR.
   Panels (b) and (e) present the impurity spectral function 
   $A(\omega)$ for Edge and Bulk locations, respectively. Sharp Fermi-level resonance peaks are observed in all cases, indicating strong Kondo resonances, between the Hubbard peaks at $\omega \simeq \pm U/2 = \pm 0.005$.  Panels (c) and (f) show the corresponding magnetic susceptibility curves ($=T\chi(T)$) as a function of temperature $T$. The Bulk location is nearly unaffected by strain at this chemical potential.  In contrast, the Edge location shifts the characteristic Kondo temperature to higher values as strain increases.  
   (g) $T_K$ as a function of strain for both $p_3$ locations: blue curve for Bulk, and red curve for Edge, show $T_K$ enhancement by over an order of magnitude at the Edge location.} 
    \label{fig4} 
\end{figure}

\end{widetext}

\subsection{Hollow site configuration - $p_3$}

\subsubsection{Effect of strain for different locations on ribbon}

Figure \ref{fig4} shows representative hybridization function curves $\Delta(\omega)$ for an impurity in a hollow site configuration $p_3$ (located at two different positions on the ZKR) for different values of applied strain. In panel (a), the impurity is on a hollow site right at one zigzag edge of the ribbon (labeled `Edge'), while in panel (b) the hollow site is in the middle of the ribbon (`Bulk'). The ZKR Fermi energy is near the upper VHS ($\mu =-0.40$), and the strain is applied at $\theta=40^\circ$.
Notice that increasing strain moves the $\Delta(\omega)$-features (red and green curves) associated with the edge of the subbands closer to the Fermi energy--the peak at $\omega \approx 0.02$ (no strain, blue curve) comes from the weakly dispersive edge-state in the ribbon, while the peak at $\omega \simeq 0.08$ corresponds to a subband that extends throughout the ribbon.  

An essential difference between the Edge/Bulk locations for hollow configurations is that $\Delta(\omega)$ is much larger overall at the Edge site, leading to stronger impurity screening even when the configurations are locally identical.  This is evidenced by the resulting impurity spectral functions $A(\omega)$, shown in panels \ref{fig4}(a) and (d), respectively. The impurity spectral densities $A(\omega)$ exhibit a sharp resonance at the Fermi level, which is characteristic of robust Kondo screening of the impurity, in addition to Hubbard peaks near $\omega =\pm U/2$. Notice that $A(0)$ values are inversely proportional to the corresponding hybridization weight $\Delta(\omega)$, as expected from Fermi-liquid theory \cite{hewson1993kondo}. It is interesting that the strong asymmetry of the hybridization function is clearly reflected in the asymmetry of $A(\omega)$ for the Edge location, while it is negligible in the Bulk.  In fact, for the largest strain values (green curves) for the Edge location, the Hubbard peaks are quite asymmetrically placed around the Fermi level, and non-universal features appear, such as the excitation at $\omega \simeq 0.008$, associated with the additional edge-state structure in $\Delta$ that has been shifted to lower excitation by the strain.  Similar non-universal features have been seen and analyzed in other cases with non-monotonic hybridization functions \cite{almeida2025identifying,PatriciaCNT}. 

To better characterize the impurity screening in these configurations, we calculate the impurity magnetic susceptibility and show $T\chi(T)$ as a 
function of temperature in panels (c) and (f). In both cases, the screening becomes more effective as the temperature drops, suppressing $\chi$ and eventually reaching the Kondo strong-coupling fixed point. The characteristic Kondo scale using the Wilson criterion, $T_K \,\chi(T_K)=0.0701$ \cite{krishna1980renormalization}, shows that the strain has nearly negligible effect for the Bulk location in Fig.\ \ref{fig4}(f).  However, $T_K$ grows by nearly an order of magnitude for the hollow site at the Edge position. Figure \ref{fig4}(g) presents a direct comparison of the Kondo temperature as a function of strain for these two configurations. 

Note that the stark difference in $T_K$ behavior with strain depends not only on the Edge vs Bulk location, but on the Fermi energy being close to high-value features in the hybridization function.  As we will see below, changing the chemical potential plays also a crucial role in the screening response.

\begin{figure}[ht]
    \centering
    \includegraphics[width=1.0\columnwidth]{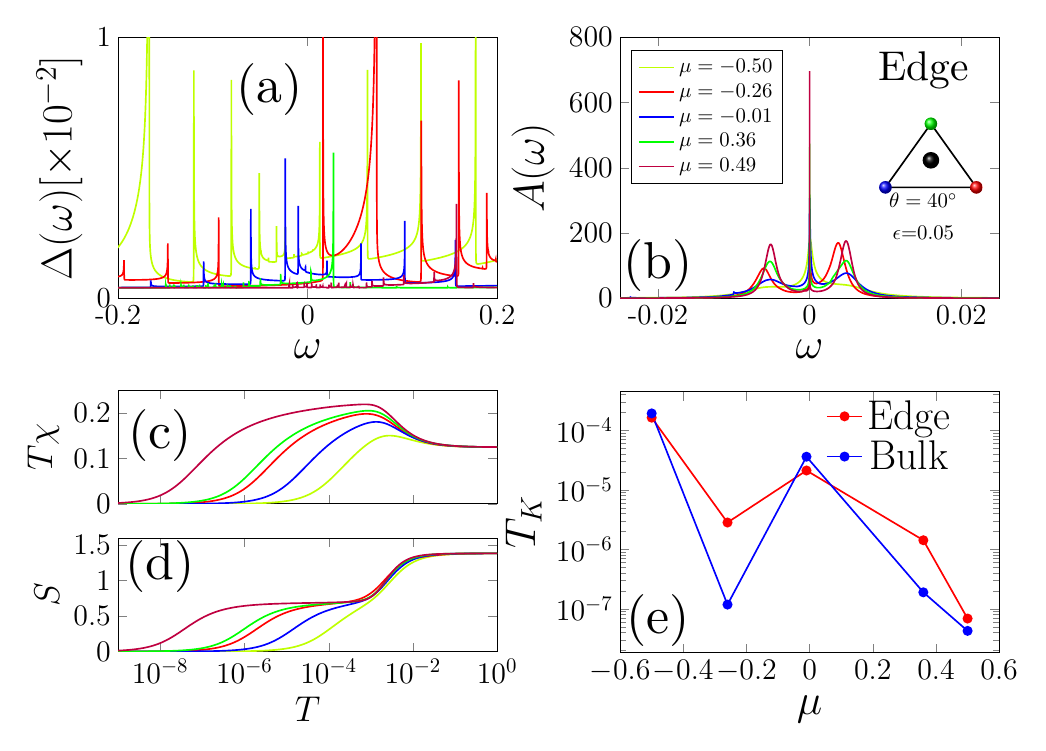}
    \caption{Impurity screening in $p_3$ configuration located on the Edge of the ribbon for different chemical potential values. The ribbon is under uniaxial strain $\epsilon=5\%$ at $\theta=40^\circ$.  (a) Hybridization function $\Delta(\omega)$ for different $\mu$ values, chosen to align $E_F$ with different band features: $\mu=-0.50$ ($-0.01$) place $E_F$ at the higher (lower) VHS; for $\mu=-0.26$, $E_F$ is near the Dirac point; $\mu=0.36$ shifts $E_F$ to the zigzag edge state; and $\mu=0.49$ shifts $E_F$ to the flat-band region of the ribbon.  (b) Corresponding spectral functions $A(\omega)$ exhibit sharp Kondo resonance near/at $E_F$, Hubbard bands, as well as non-universal features, especially for $\mu=0.49$ and $-0.01$, where sharp features in 
    $\Delta(\omega)$ values are close to $E_F$. (c)  Magnetic susceptibility $T\chi(T)$ shows the evolution from the free orbital fixed point at high $T$, to the local moment at intermediate  $T$, and finally strong coupling at low $T<T_K$. The temperature dependence of the entropy $S$ shows similar evolution. 
    (d) Kondo temperature $T_K$ as function of $\mu$. The blue curve corresponds to results near the zigzag edge (Edge), while the red curve represents response in the middle of the ribbon (Bulk). Lines drawn to guide the eye.} 
    \label{fig5} 
\end{figure}

\subsubsection{Chemical potential dependence of impurity screening}

To illustrate the role of chemical potential tuning on impurity screening, Fig.\ \ref{fig5} shows results for the effective hybridization $\Delta(\omega)$, spectral function $A(\omega)$, entropy $S(T)$ and magnetic susceptibility, for an impurity with $p_3$ hollow site hybridization on a ZKR Edge location, for different values of chemical potential.  The analysis of these quantities allows one to understand the system's ground state under various conditions.
The chemical potential values were selected so that the Fermi energy 
is close to the characteristic features of the ZKR DOS, and probed through the hybridization function. For $\mu = -0.5$, the Fermi level $E_F$ is located near the upper VHS; $\mu = -0.26$ shifts $E_F$ close to the Dirac point; $\mu = -0.01$ corresponds to $E_F$ near the lower VHS; $\mu = 0.36$ places $E_F$ near the energy of the zigzag edge state, and finally, $\mu = 0.49$ shifts $E_F$ near the flat-band at the bottom of the ZKR DOS. 
Looking at panel (a), it is evident that changes in the chemical potential result in significant changes on the value of $\Delta(\omega \simeq 0)$, showing various subband features near the Fermi level.  For example, despite the fact that $\mu = 0.49$ places $E_F$ near the ribbon's flat-band and its associated high DOS, the corresponding susceptibility in panel \ref{fig5}(c) (dark red curve) shows a strongly suppressed $T_K$.  This suppression can be traced back to the fact that in this Edge location, the $p_3$ hollow site has minimal contribution from the flat-band states, since they have little weight near the edges (see Fig.\ \ref{fig3}). This makes the hybridization function for this Edge configuration to have a much lower amplitude, as seen in \ref{fig5}(a). 

For $\mu = 0.36$ (dark green curves), $E_F$ lies near the energy of the zigzag edge state, which results in a larger $\Delta(0)$. This enhancement in turn drives a substantial rise in $T_K$, as seen in Fig.~\ref{fig5}(c). 
We should also notice that when $E_F$ lies near one of the VHSs, the Kondo temperature is even more substantially enhanced, as exemplified by the results for $\mu =-0.5$ and $-0.01$, with the highest $T_K$ values. 
We should point out that the resulting Kondo resonance and associated Hubbard bands seen in the spectral functions in Fig.~\ref{fig5}(b) may also exhibit non-universal spectral features. This is due to the proximity of highly non-monotonic features in the different $\Delta(\omega)$, and perhaps most noticeable for $\mu=-0.5$ ($E_F$ near the high energy VHS, light green curves), where the hybridization is so strong that the local moment fixed point is barely defined--see Fig.~\ref{fig5}(c), where no plateau is visible in $S$ at this value--and the weak Hubbard bands suggest that the system is near the mixed valence regime.

These results demonstrate that fine-tuning the chemical potential to scan prominent features in the DOS can result in large changes in $T_K$, even when the impurity location and local configuration ($p_3$ in this case) remain unchanged.  
Figure \ref{fig5}(d) shows a comparison of $T_K$ values for the Edge location and a similar $p_3$ hollow configuration in the Bulk of the ribbon. Shifts in $\mu$ produce drastic changes in $T_K$ in both cases, although the variation is more substantial for the Bulk location, and results often in the highest $T_K$. Notice the largest $T_K$ shown is whenever $E_F$ is near a VHS ($\mu = -0.5$ and $0$), and the DOS is higher.

\begin{widetext}

\begin{figure}[ht]
    \centering
    \includegraphics[width=1.0\columnwidth]{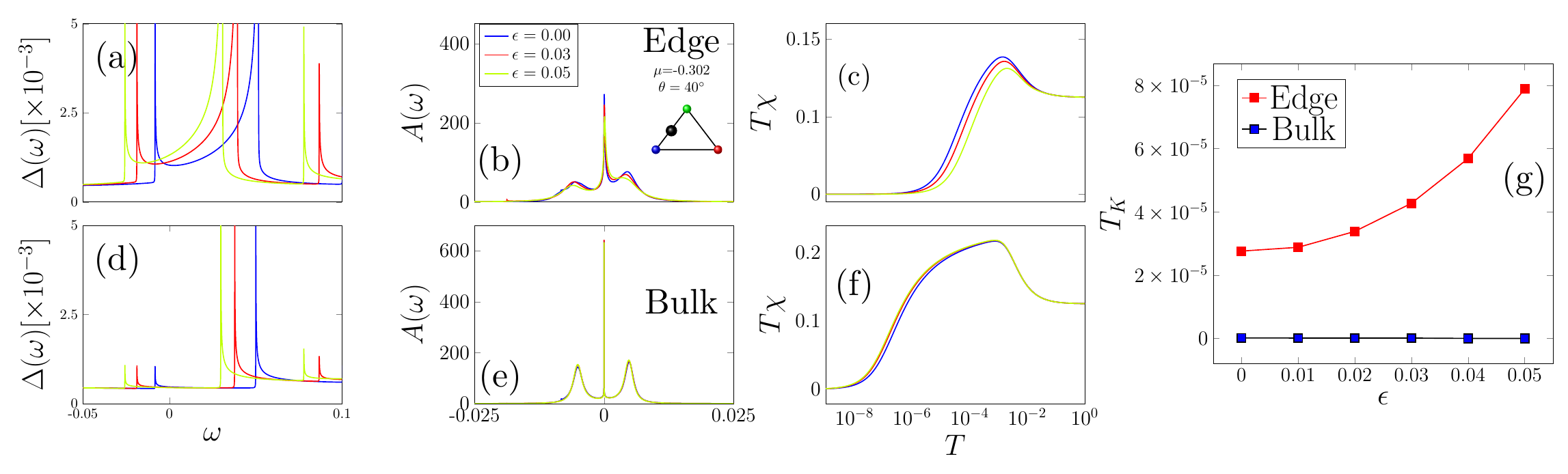}
    \caption{Results for the impurity located in ${p}_2$ bridge configurations, for two positions on the ZKR--Edge and Bulk--and different strain magnitude $\epsilon$. Here, $\mu=-0.30$, $\theta = 40^{\circ}$. Panels (a)-(d) show the hybridization function for different locations. (b)-(e) show corresponding impurity spectral functions. (c)-(f) show the magnetic susceptibility. (g) Comparison of $T_K$ for Edge and Bulk locations in bridge configuration $p_2$.} 
    \label{fig6} 
\end{figure}

\end{widetext}

\subsection{Bridge configuration - $p_2$}

\subsubsection{Effects of strain magnitude on different positions}

A different hybridization environment is the bridge configuration $p_2$, illustrated in Fig.\ \ref{fig1}(a).  Comparing results for different $p_2$ locations on the nanoribbon, Edge and Bulk, we notice significant differences in impurity screening. Figure \ref{fig6} presents results for this situation, with $\mu=-0.30$, which places  $E_F$ between the Dirac point and the higher-energy edge state; states at $E_F$ are characterized by being mostly extended throughout the ribbon. The figure shows the hybridization function, the impurity spectral function, and the magnetic susceptibility results.

As in the $p_3$ configuration, we find that $\Delta(\omega)$ is much larger for the Edge location than for the Bulk, and that sharp features associated with the ribbon edge state shift away from $E_F$ as strain increases at this $\mu$ value. The spectral functions exhibit robust Kondo resonances at the Fermi level in all cases, showing clear universal behavior in the Bulk (well-defined Kondo resonance and Hubbard bands), as $\Delta(\omega \simeq0)$ is featureless for all values of strain in this case--see Fig.\ \ref{fig6}(d).  For the Edge location, $A(\omega)$ exhibits non-universal features, including asymmetric Hubbard bands, as prominent peaks in $\Delta$ appear near the Fermi energy, resulting in particle-hole asymmetry. Notice also that  $A(0)$ decreases for larger $\Delta(0)$ values, as one would expect. The magnetic susceptibility shows low-energy shifts that indicate variations in $T_K$ as the temperature drops, reflecting the system traversing different fixed points.  

A compilation of $T_K$ results is shown in Fig.\ \ref{fig6}(g), demonstrating weak sensitivity to strain when in the Bulk location, and a higher sensitivity when the impurity is on the Edge of the nanoribbon in the $p_2$ configuration, increasing by nearly one order of magnitude. Interestingly, $T_K$ drops slightly with strain in the Bulk location.

\begin{figure}[ht]
    \centering
    \includegraphics[width=1.0\columnwidth]{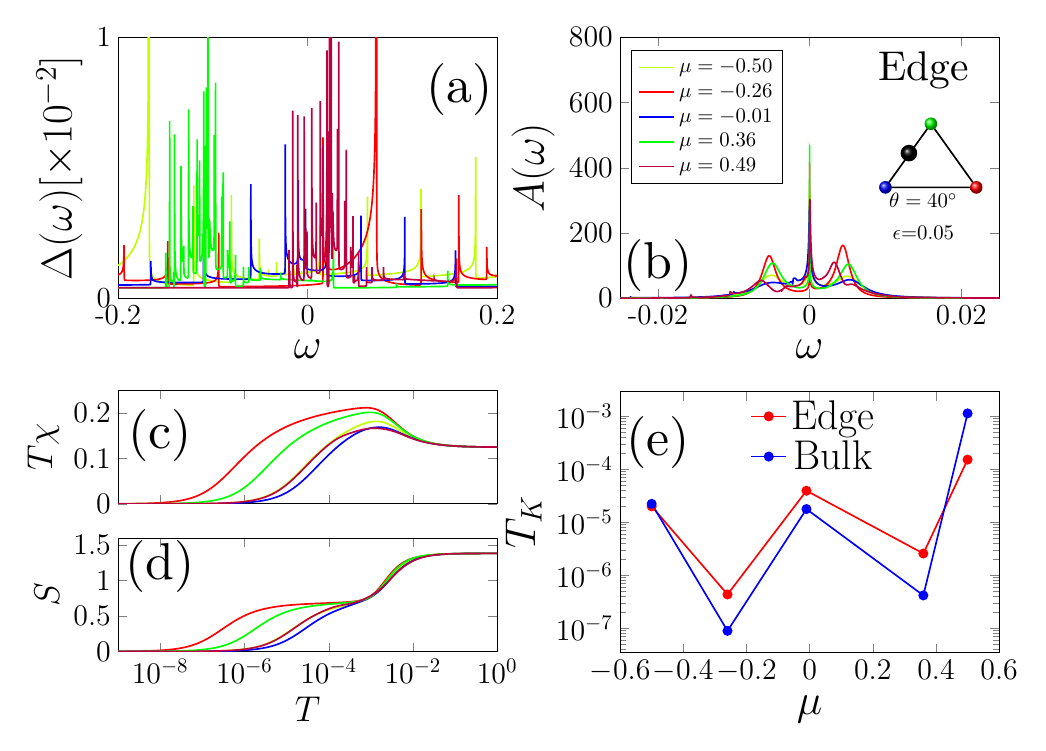}
\caption{Impurity screening with ${p}_2$ configuration on lattice sites at the Edge of the ribbon, for different chemical potential values. The ZKR is under strain $\epsilon=5\%$ at $\theta=40^\circ$. (a) Hybridization function $\Delta(\omega)$ for different $\mu$ values, chosen to align $E_F$ with different band features: $\mu=-0.50$ ($-0.01$) places $E_F$ at the higher (lower) VHS; for $\mu = -0.26$, $E_F$ is near the Dirac point; $\mu = 0.36$ shifts $E_F$ to the ribbon edge state; while $\mu  = 0.50$ shifts $E_F$ to the flat-band region of the ribbon. 
(b) Spectral functions $A(\omega)$ exhibit Kondo resonance near/at $E_F$, Hubbard bands, as well as non-universal features associated with sharp non-monotonic features in $\Delta(\omega)$  close to $E_F$. (c) Magnetic susceptibility $T \chi(T)$  and entropy $S$ show characteristic strong coupling behavior at low $T < T_K$. (d) Summary of $T_K$ for $p_2$ bridge configuration for both Edge and Bulk sites.} \label{fig7}
   \end{figure}

\subsubsection{Dependence on the chemical potential}

Figure \ref{fig7} presents results for the impurity on the bridge configuration $p_2$ and located at the Edge of the nanoribbon. We vary the chemical potential to explore its role on controlling the screening for this site-asymmetric configuration. Notice the values of $\mu$ explored are the same as in Fig.\ \ref{fig5}, so that the Fermi energy lands near important spectral features in the kagome band structure. Comparison of the hybridization function $\Delta(\omega)$ for $p_3$ and $p_2$ configurations--Fig.\ \ref{fig5}(a) vs \ref{fig7}(a)--for the Edge location shows notable differences, especially for $\mu=0.36$ and $0.49$.  The hollow site configuration results in overall smaller values of $\Delta(0)$, and strongly suppressed features at the onset of various subbands in the structure.  This contrasts with the bridge configuration, which shows much larger $\Delta$ values and arises from destructive interference between the three atomic sites in a $p_3$ configuration. The interference is clearly constructive in the $p_2$ configuration. As a consequence, the impurity screening shows different behavior in the two configurations, as evident in their respective spectral functions and magnetic susceptibilities. 

A summary of $T_K$ values for the $p_2$ configuration is shown in Fig.\ \ref{fig7}(d), which also includes results for the Bulk location. $T_K$ has a non-monotonic behavior with $\mu$, as anticipated, while the Bulk values are consistently smaller here.  A larger $T_K$ is an exception for $\mu=0.49$, when the Fermi energy lies near the kagome flat-band, demonstrating the strong spectral weight of such states in the middle of the ribbon.

\subsection{On top configuration - $p_1$}

\subsubsection{Effects of strain magnitude}

\begin{widetext}

\begin{figure}[ht]
    \centering
    \includegraphics[width=1.0\columnwidth]{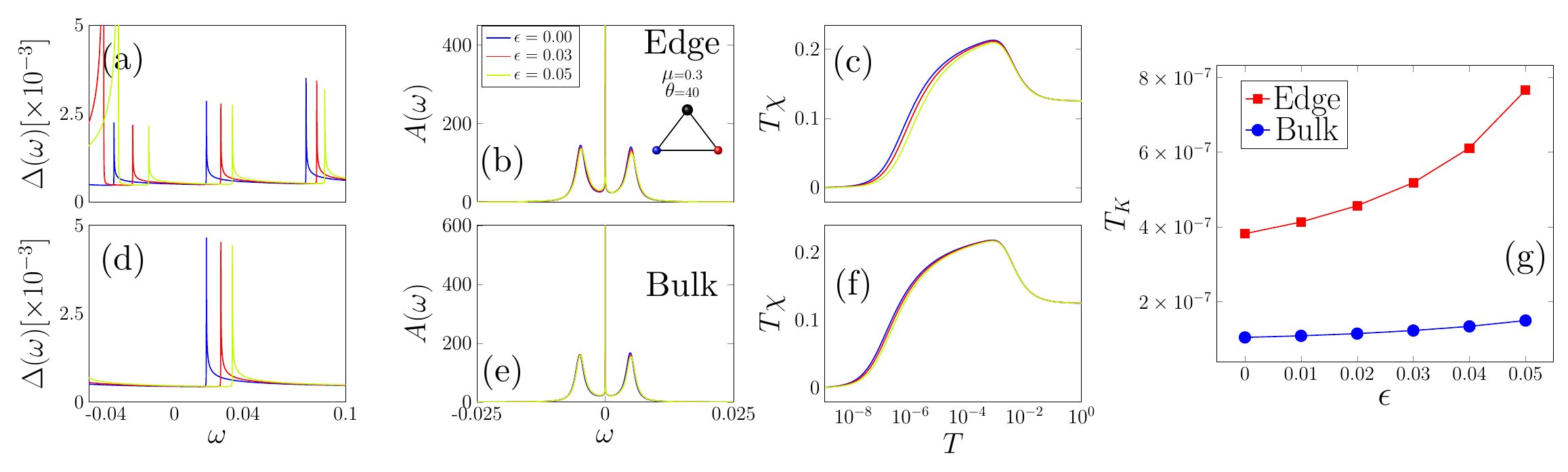}
    \caption{Results for impurity with on-top configuration ${p}_1$ at two different positions, Edge and Bulk, and different strain values. Here, $\mu=0.30$ and $\theta = 40^{\circ}$. Panels (a) and (d) show the hybridization function for Edge and Bulk positions, respectively. (b) \& (e) Impurity spectral functions show a clear Kondo resonance and associated Hubbard bands. (c) \& (f) Magnetic susceptibility. (g) Kondo temperature comparison for the two on-top, $p_1$, locations vs strain.
    } 
    \label{fig-p1}
\end{figure}

\end{widetext}

Figure \ref{fig-p1} shows the hybridization function for an impurity on top of a lattice site, $p_1$ case, in addition to the impurity properties (spectral function and 
magnetic susceptibility) for different strain values and at two different locations on the ribbon, Edge and Bulk. In this case, we have chosen $\mu=0.30$, placing the Fermi energy in a relatively smooth region of the DOS.
Notice that $\Delta(\omega)$ in Fig.\ \ref{fig-p1}(a) shows sharp features shifting away from $E_F$ as strain increases, while $\Delta(0)$ changes little with strain.  This results in the screening response being weakly dependent on the applied strain in both Edge and Bulk locations. 
Consequently, the spectral function and susceptibility exhibit nearly standard universal behavior. $A(\omega)$ shows well-defined Hubbard peaks and a sharp Kondo resonance. The susceptibility shows a standard temperature dependence, visiting the appropriate fixed points of the problem.

Figure \ref{fig-p1}(g) presents a comparison of the Kondo temperature as a function of strain amplitude for the on-top configuration $p_1$ for Edge and Bulk locations on the ribbon. 
As expected, the Edge location exhibits $T_K$ values an order of magnitude larger than in the Bulk, while strain increases $T_K$ by only a factor of two in both cases.

\begin{figure}[ht]
    \centering
    \includegraphics[width=1.0\columnwidth]{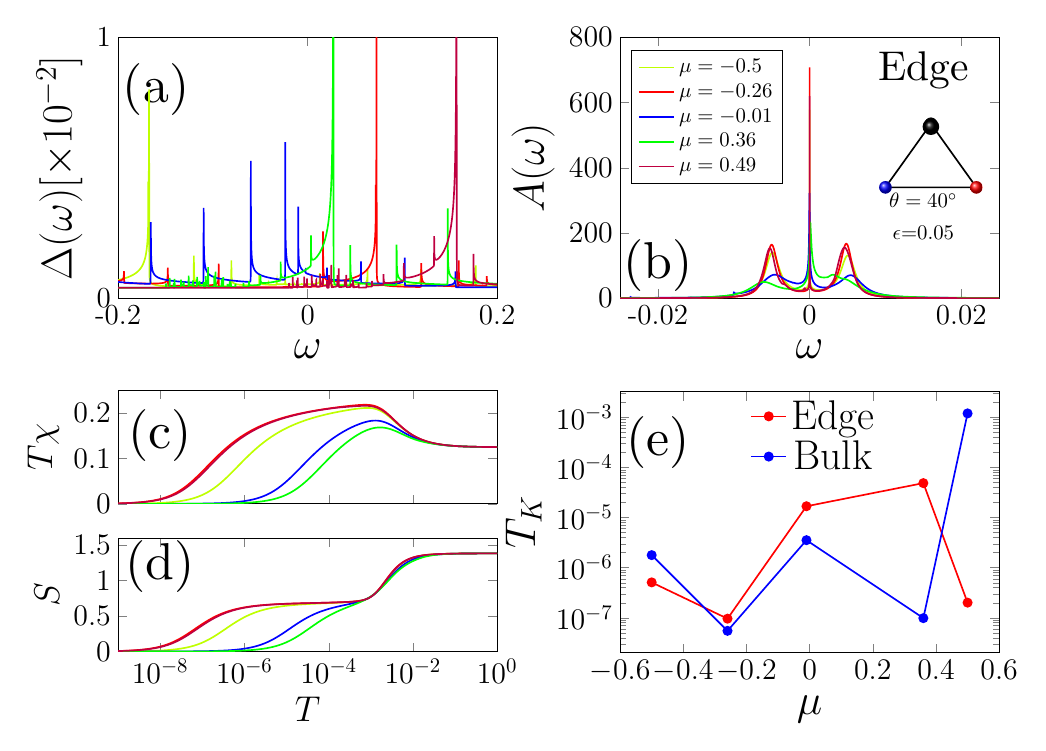}
    \caption{Impurity results for on-top configuration ${p}_1$ at Edge location for different values of the chemical potential $\mu$.  Uniaxial strain $\epsilon=0.05$ and $\theta = 40^{\circ}$.  (a) The hybridization function is rather similar to that in Fig.\ \ref{fig5}.
    (b) Impurity spectral function shows universal Kondo behavior, including a sharp resonance at the Fermi level and well-defined Hubbard bands.  Non-universal features are seen for $\mu=0.36$. (c) Corresponding magnetic susceptibility $\chi(T)$ and entropy $S$ as a function of temperature $T$. (d) Evolution of the Kondo temperature $T_K$ as a function of $\mu$ for $p_1$ configuration at both Edge and Bulk locations. 
    } 
    \label{fig-p12} 
\end{figure}

\subsubsection{Dependence on chemical potential}

Figure \ref{fig-p12} presents the hybridization,  
impurity spectral function, magnetic susceptibility and entropy corresponding to an impurity with an on-top configuration ${p}_1$ and placed at the Edge of the ZKR.  We vary the chemical potential, while keeping strain $\epsilon = 0.05$ and $\theta = 40^\circ$ fixed. The chosen values of $\mu$ are the same as in Fig.\ \ref{fig5} and \ref{fig7}, allowing us to illustrate the effect of the different hybridization environment.
Comparison of the $\Delta(\omega)$ function in this figure with the corresponding one in Fig.\ \ref{fig5}(a) shows similar features and overall amplitude (with subtle variations under close inspection), and larger differences with that in \ref{fig7}(a), again emphasizing the importance of the different hybridization geometry. The spectral functions in \ref{fig-p12}(b) show well-defined Kondo resonance and Hubbard bands.  An exception is seen for $\mu=0.36$, which places the Fermi level near the ribbon edge state, where it naturally exhibits a much stronger $\Delta(0)$, together with prominent sharp features for $\omega\simeq 0$ (bright green curve).  The strong hybridization results in a broader Kondo resonance (and a larger $T_K$), as well as poorly defined Hubbard bands, because it pushes the impurity closer to a mixed-valence regime. 
As in other configurations, the resulting Kondo temperature shows non-monotonic behavior as $\mu$ varies, as seen in panel \ref{fig-p12}(e).  For this $p_1$ geometry, there is a marked contrast between the Edge and Bulk locations for the two positive values of $\mu$ shown, where they have opposite $T_K$ values.

\section{Conclusions}

In summary, we have investigated the onset and characteristics of the Kondo effect in zigzag kagome ribbons under uniform uniaxial strain, analyzing how magnetic impurity screening varies across different system parameters, bonding configurations, and ribbon locations. Specifically, we explored the impact on the Kondo temperature when modifying the impurity's position within the ribbon, as well as the effect of shifting the chemical potential. The rich features of the kagome lattice, inherited by the nanoribbon, provide a host system in which to explore the role of lattice geometry and associated effective hybridization in screening a magnetic impurity. We find that the Kondo effect and its associated Kondo temperature vary strongly with the impurity's location on the ribbon, providing an interesting probe of the local orbital environment.  
Ideally, one could consider monitoring $T_K$ values across different but identical impurities, as they experience different orbital environments.  These could be modiffied by local gating or built-in strains.  Such a $T_K$ map would provide an interesting perspective on the role of local hybridization environments.  This would be especially true if it could be combined with powerful scanning tunneling microscopy techniques for atomic manipulation.
The system could, in turn, be further modified by applied strain to help explore the role of local environments on impurity screening.

We have found substantial increases in the Kondo temperature under applied strain, most notably when the impurity is positioned near the ribbon’s edge. This increase may span several orders of magnitude, suggesting that, for a more easily characterized investigation of the Kondo effect in zigzag kagome ribbons, the most favorable regions are near the edges. The optimal configuration corresponds to the impurity hybridized on top of an edge atom ($\vec{p}_1$ configuration), as this results in overall larger $T_K$. This would allow better contrasts to explore the effects of strain on the microscopic charge redistribution induced by the strain field. 

\begin{acknowledgments}
P.A.A., S.E.U. and G.B.M. acknowledge support from 
the CAPES-PrInt/UFU program. P.A.A. thanks the Brazilian funding agency CAPES 
for financial support. S.E.U. acknowledges support from the US Department of Energy, 
Office of Basic Energy Sciences, Materials Science and Engineering Division.

\end{acknowledgments}

\bibliography{refS}

\end{document}